\documentstyle[11pt,newpasp,twoside,epsf]{article}
\def\edcomment#1{\iffalse\marginpar{\raggedright\sl#1\/}\else\relax\fi}
\reversemarginpar
\begin{document}
\title{Molecular Gas in The Central Kpc of Starbursts and AGN}

\author{Shardha Jogee\altaffilmark{1}, Andrew J. Baker\altaffilmark{2}, 
Kazushi Sakamoto\altaffilmark{3}, Nick Z. Scoville\altaffilmark{1}, 
and Jeffrey D. P. Kenney\altaffilmark{4}}
\altaffiltext{1}{California Institute of Technology, MS 105-24, Pasadena, CA 91125}
\altaffiltext{2}{MPI f{\" u}r extraterrestrische Physik, Postfach 1312, 85741 Garching,Germany}
\altaffiltext{3}{Harvard-Smithsonian Center for Astrophysics, 
SMA,  P. O. Box 824, Hilo, HI 96721}
\altaffiltext{4}{Yale University Astronomy Department, New Haven, CT 06520-8101}


\begin{abstract}
With the recent advent of large interferometric surveys, 
we can now probe the physical conditions, dynamics, 
and  star-forming  properties  of molecular gas in the 
central kpc of starbursts and active galactic nuclei. 
We present results from the high-resolution ($\sim$ 100 pc) 
interferometric  survey  of molecular 
gas in the inner kpc of nearby starbursts (Jogee, 
Kenney, \& Scoville 2001a)  
and  the  ongoing  multi-transition  
survey of cold, warm, and dense molecular gas in a broad range of 
active and inactive galactic nuclei 
(Jogee, Baker, Sakamoto, \& Scoville 2001b). 
\end{abstract}

\section{Survey of molecular gas in the inner kpc of starbursts}

Using the Owens Valley Radio Observatory (OVRO) in the past five years,  
a high resolution ($\sim$ 100 pc) interferometric CO  (J=1-$>$0)  survey of 
molecular gas  in the inner kpc of eleven circumnuclear starbursts and  
non-starbursts has been conducted (Jogee 1999; Jogee et al. 2001a) 
The sample  shows an order of magnitude variation in circumnuclear 
molecular  gas content and  star formation efficiency (SFE),  defined 
as the star formation rate per  unit mass of molecular gas.  
The  circumnuclear starbursts, characterized by a high SFE, 
include the brightest nearby  starbursts comparable to M82. 
The survey investigates the physical conditions, dynamics, 
and  star-forming  properties  of molecular gas in the inner kpc. 
Selected results are outlined in $\S$ 1.1-1.2.

\subsection{The molecular environment}

Molecular gas in the inner  kpc differs markedly from 
the  the outer disk (see Table 1). 
In the inner kpc, gravitational instabilities can only be triggered at high 
gas densities (few 100-1000 M$_{\tiny \odot}$ pc$^{-2}$) due to the 
large Coriolis and pressure forces resulting from the 
large epicyclic frequency  
and velocity dispersion.  
Gravitational instabilities  can promote star formation  by 
agglomerating  molecular clouds into complexes where they can grow 
by collision, coalescence, and accretion. 
In the inner kpc, the growth  timescale  ($t_{\rm GI}$)  of 
gravitational instabilities can be  so  short (a few Myrs) 
that it is comparable to the lifetime  of  OB stars which destroy 
molecular clouds. One may therefore expect the fraction of gas converted 
into stars before cloud disruption to be higher 
in circumnuclear starbursts than in the outer disk. Detailed
studies are required to further investigate this possibility.  
%
Another special feature of the inner kpc  is the presence of a 
high pressure, high turbulence molecular ISM. 
Such an ISM  is believed to favor   the formation of more 
massive clusters  (e.g., Elmegreen 1993) and may explain why 
bright super star clusters in non-interacting spirals 
tend to occur preferentially in the  inner kpc.   
Table 1 also shows that 
the prototypical ultra luminous infrared galaxy (ULIRG) Arp 220  
looks like a scaled-up  version of the nearby circumnuclear 
starbursts.  This raises  the possibility that  ULIRGs may be 
starbursts which have built an extreme  molecular environment 
(density and linewidths) in the central part of deep  potential well, 
through major mergers or interactions.

\begin{table}[t]
\caption{Molecular gas properties in the inner kpc}
\begin{center}
\renewcommand{\arraystretch}{1.4}
\setlength\tabcolsep{5pt}
\begin{tabular}{lccc}
\tableline
\vspace{-0.15cm}
 {Quantities   }   &  {Outer Disk}      &
 {Inner r=500 pc}      &  {Inner r=500 pc} \\
 {}   &  { of Sa-Sc}      &
 {of sample starburst}          &  {of Arp 220}\\ 
\tableline
(1) $M_{\rm gas-m}$ [$M_{\tiny \odot}$] & 
$\le$ few $\times 10^{9}$ & 
Few $ \times$ (10$^{8}$-10$^{9}$)   & 
$ 3 \times 10^{9}$     \\
(2) $M_{\rm gas}$/$M_{\rm dyn}$  [\%] & 
$<$ 5    & 
10 to 30      &  
40 to 80      \\
(3) SFR [$M_{\tiny \odot}$ yr$^{-1}$] & -  & 
0.1-11 
&  $>$ 100  \\
(4) $\Sigma_{\rm gas,m}$ [$M_{\tiny \odot}$ pc$^{-2}$] & 
1-100 & 
500-3500    &  
$ 4 \times 10^{4}$   \\
(5)  $\sigma$  [km s$^{-1}$] & 
6-10 & 
10-40
& 90  \\
(6) $\kappa$ [km s$^{-1}$  kpc$^{-1}$] & 
$<$ 100 & 
800-3000   &  
$>$ 1000   \\
(7) $\Sigma_{\rm crit}$ [$M_{\tiny \odot}$ pc$^{-2}$]& 
$<$ 10   & 500-1500   & $>$ 2000  \\
(8) $t_{\tiny \rm GI}$ [Myr]& 
$>$10    & 0.5-1.5    & $<$ 1   \\
(9) $\lambda_{\tiny \rm J}$ [pc&
Few $\times$ (100-1000)    & 100-300   & 90  \\
\tableline
\tableline
\end{tabular}
\end{center}
\label{apptab1b}
\small
~ Rows are : 
(1) $M_{\rm gas-m}$,  the molecular gas mass
(2) $M_{\rm gas,m}$/M$_{\rm dyn}$, the ratio of  molecular gas mass 
to dynamical mass;
(3) SFR, the star formation rate; 
(4) $\Sigma_{\rm gas,m}$ the 
surface density;
(5) $\sigma$,  the 
velocity dispersion; 
(6) $\Sigma_{\rm crit}$, the critical 
density for the onset of gravitational instabilities
(7) $\kappa$,  the epicyclic frequency; 
(8) $t_{\tiny \rm GI}$ = Q/$\kappa$, the growth  
timescale of the most unstable wavelength 
(9) $\lambda_{\tiny \rm J}$, the Jeans length 
\normalsize
\end{table}

\subsection{Molecular gas distribution and triggers of star formation}

The  molecular gas shows a wide variety of  morphologies 
(Fig.~1) ranging from   relatively axisymmetric annuli or disks 
(NGC\,4102, NGC\,3504, NGC\,4536,   and  NGC\,4314), 
elongated double-peaked  and spiral morphologies (NGC\,2782, NGC 3351, and  
NGC 6951)  to extended distributions elongated along the 
large-scale bar (NGC\,4569).
We find  large gas concentrations inside  the outer inner Lindblad 
resonance (ILR)  of the bar, consistent with theory (e.g., Combes \& Gerin 1985). 
Assuming the  epicycle theory for a weak bar, it is estimated that 
in the sample galaxies, typically 
the bar pattern speed is $> 40$--115\,km~s$^{-1}$~kpc$^{-1}$, 
the  radius of the outer ILR is  $>$\,500\,pc, 
and the radius of the inner ILR is  $<$\,300\,pc. 
(Jogee et al.  2001a).

\begin{figure}[h]
\plotfiddle{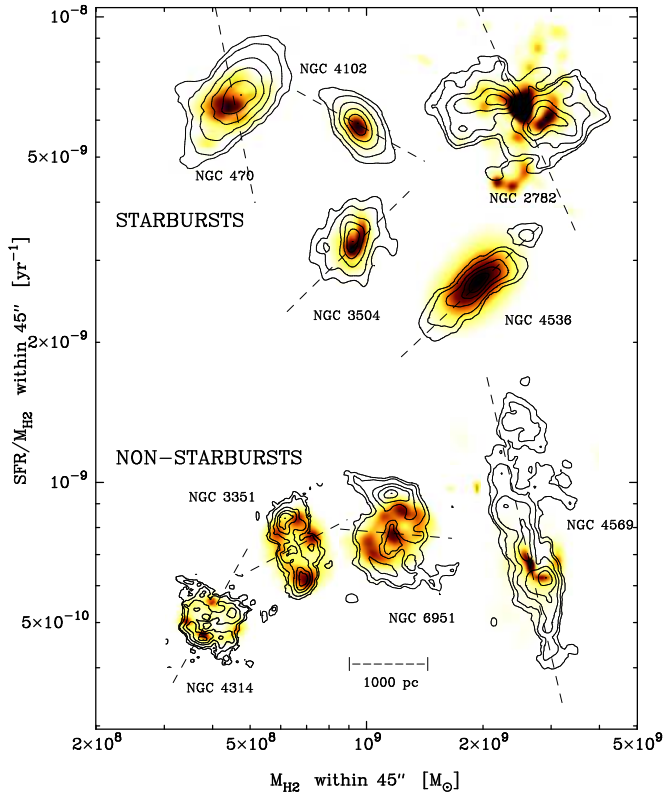}{2.5in}{0}{45}{45}{-150}{-180}
\vspace{2.0in}
\caption[]{
In the SFR/M$_{\mathrm{H2}}$ vs.~M$_{\mathrm{H2}}$ plane, 
the CO  intensity  (contours) is  overlaid on the  
star formation  (greyscale), as traced by RC  and H$\alpha$. 
The dotted line is  the P.A. of the  large-scale stellar bar/oval. 
The synthesized CO beam is 100--200\,pc.
}
\end{figure}

The starbursts and  non-starbursts have  circumnuclear SFR 
of  3--11   and  0.1--2\,M$_\odot$\,yr$^{-1}$, respectively. 
For a given  CO-to-H$_{\rm 2}$ conversion factor,
the starbursts have a larger peak gas surface density $\Sigma_{\rm gas-m}$ 
in the inner 500\,pc radius than non-starbursts  
with a similar circumnuclear gas content (Fig.~2a).  
In the regions of intense star formation, $\Sigma_{\rm gas-m}$  
remains close to the Toomre (1964) critical density ($\Sigma_{\rm crit}$) for 
the onset of gravitational instabilities, despite 
an order of magnitude variation in $\Sigma_{\rm crit}$  (Fig.~3b and e). 
In the non-starbursts, there are gas-rich regions with no 
appreciable  star formation, for instance, at the CO peaks in NGC\,6951 
and inside the ring of HII regions in NGC\,3351 and NGC\,4314.
The gas surface density, although high,  is still 
sub-critical in regions of inhibited star formation,
as illustrated for NGC\,4314 in Fig.~3e--f.  

\begin{figure}[t]
\plotfiddle{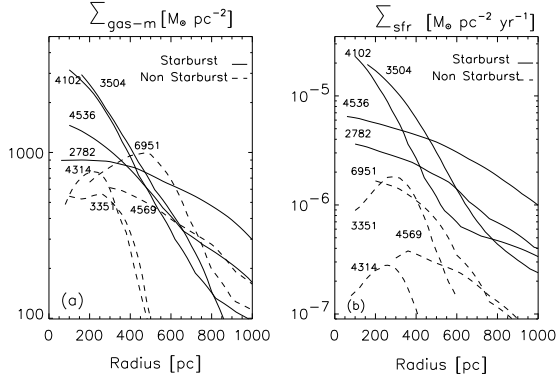}{1.0in}{0}{38}{38}{-150}{-160}
\vspace{0.65in}
\caption
{\bf(a) \rm The azimuthally-averaged  molecular gas 
surface density  
and 
\bf (b) \rm 
the extinction-corrected SFR per unit area.
}
\end{figure}

\begin{figure}[h]
\plotfiddle{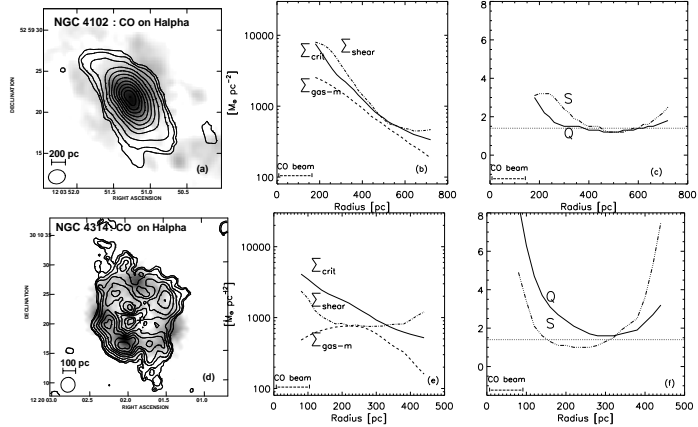}{2.0in}{0}{55}{55}{-180}{-130}
\vspace{-0.1in}
\caption[]{
\rm
{\bf (a, d)} CO   (contours) on 
 H$\alpha$ (greyscale) distributions.  {\bf (b, e)} $\Sigma_{\rm gas-m}$, 
$\Sigma_{\mathrm{crit}}$, and  $\Sigma_{\mathrm{shear}}$. 
{\bf (c, f)} Toomre Q and shear S parameters (Jogee et al 2001a). 
Quantities are plotted starting at a radius equal to the 
CO beam  size ($\sim$2$''$). 
See text for details. 
}
\end{figure}

\section{Survey of cold, warm, and dense  gas in active and inactive nuclei}

\begin{figure}[h]
\plotfiddle{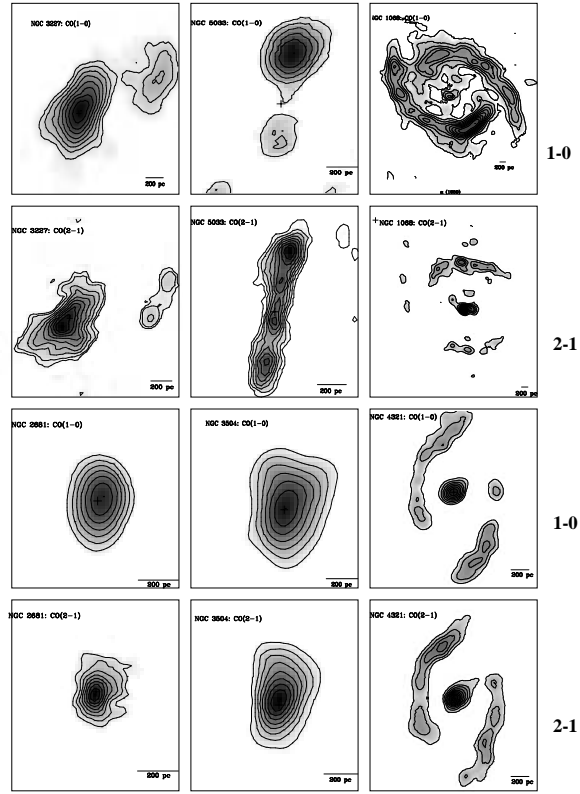}{1.5in}{0}{50}{50}{-170}{-260}
\vspace{2.7in}
\caption
{CO(1--0) and CO(2--1) emission  in the inner kpc of select 
AGN (NGC 3227, NGC 1068, NGC 2681, and  NGC 5033) and IGN (NGC 3504 and NGC 4321) 
}
\end{figure}

\begin{figure}[h]
\plotfiddle{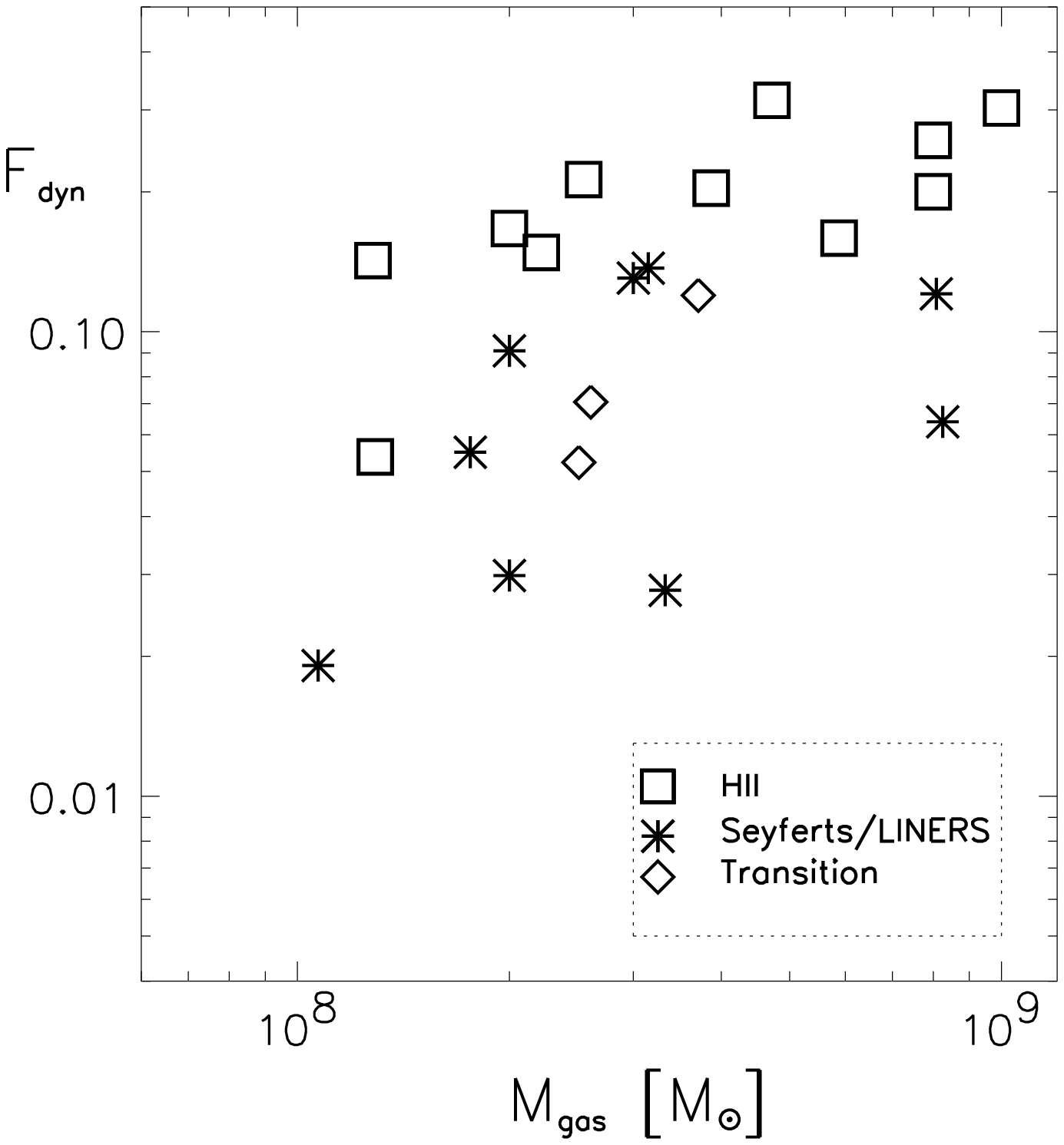}{1.0in}{0}{30}{30}{-150}{-90}
\vspace{0.5in}
\caption
{Preliminary plot  of 
the molecular gas mass  fraction $F_{\rm dyn}$  and mass $M_{\rm gas}$  
in the inner 500 pc.  
}
\end{figure}

We present  the ongoing  multi-transition  high-resolution  ($\sim$ 100 pc) 
interferometric survey of cold, warm, and dense molecular gas in a broad 
range of 
nearby active (AGN) and inactive (IGN) galactic nuclei by 
Jogee, Baker, Sakamoto, and  Scoville. 
The sample consists of AGN and IGN 
whose optical types encompass  Seyfert, LINER,  and  HII-regions,  
and whose star formation efficiency varies by  an order of magnitude. 
The survey aims at exploring the physical conditions and  
dynamics of the molecular gas in AGN and IGN, and constraining 
the drivers of activity levels in galactic nuclei. 
When completed, the survey will cover  forty galactic nuclei with  
high resolution  CO(1--0)  observations and fifteen nuclei with 
multiple line  observations,  thus constituting  one of the 
largest multi-transition interferometric surveys of galactic 
nuclei to date. 
It will complement ongoing  multiple line surveys 
of  nuclei  with the  Plateau de Bure  (NuGA; Garcia-Burillo 
\& Combes, private communication)  and  Nobeyama Radio Observatory
(Kohno et al. 2001, these proceedings) interferometers.
We built the survey database by pooling together 
new OVRO  CO (1--0), C0(2--1), and  HCN(1--0) observations 
with  existing data from three high resolution surveys: the Jogee (1999) CO(1--0)  
survey of starbursts and non-starbursts, the Sakamoto et al. (1999)  
CO(1--0) survey of CO-luminous spirals, and the Baker (2000) CO(1--0) 
and CO (2--1) survey of AGN with broad  H$\alpha$ emission.

We are investigating the 
physical properties of the cold, warm, and dense molecular gas 
in AGN and IGN  with different levels of star formation, 
nuclear activity, and  radiation fields. 
Fig. 4  shows  CO(2-1) and CO(1-0) emission in the inner kpc of several AGN 
and IGN in our sample. The distributions of the cold and warm gas 
are similar in several HII  nuclei (e.g., NGC 3504, NGC 4321),  but several 
AGN  (e.g., NGC 5033, NGC 2681, NGC 1068)  show bright  or enhanced 
CO (2-1) emission near the central engine. 
Baker (2000)  found that the integrated intensity ratio of CO(2-1)  to 
CO(1-0) in temperature units  exceeds one in several AGN  
(e.g., NGC 1068, NGC 2681).  Such high ratios can result from optically thick 
gas in  a two-zone model  where  CO(2-1) comes from warm outer layers 
of externally heated clouds. By comparing AGN and IGN in our sample, 
we are investigating whether hard X-rays constitute the dominant heating 
mechanism in AGN.  We also hope to constrain the  CO-to-${\rm H_2}$ conversion 
factor by comparing  our line ratios with those from 
surveys of the Galaxy.

We are  also testing if  nuclear types are correlated with the molecular 
gas mass fraction in the inner 500 pc.
Despite mounting evidence that most galaxies contain a black hole, 
a large fraction of  galaxies do not 
show evidence for an  AGN, as characterized by 
high-excitation optical lines produced by a hard accretion-disk spectrum. 
One hypothesis is that in some systems, 
a high circumnuclear gas mass fraction  ($F_{\rm dyn}$) 
can trigger efficient circumnuclear star formation which starves a black hole 
or washes out its accretion spectrum.  
The low  Q  in circumnuclear  regions with high SFE 
(e.g., Jogee 1999), 
and  the low $F_{\rm dyn}$ 
in several AGN (e.g., Sakamoto et al. 1999) lend some support to this hypothesis. 
Preliminary results from the half of the survey data  (Fig. 5) 
suggest the HII nuclei have larger  gas mass fractions 
than the Seyfert and LINERS. We emphasize that the results are 
preliminary and an ongoing  analysis of the entire sample 
will further investigate this apparent correlation.

%


\begin{references}
\reference
Baker, A. J. 2000, Ph.D. thesis,  California Institute of Technology

\reference
Combes, F., \& Gerin, M. 1985, \aap, 150, 327


\reference
Elmegreen, B. G. 1993, ApJ, 411, 170




\reference
Jogee, S. 1999, Ph.D. thesis,  Yale University

\reference
Jogee, S.,  Kenney, J. D. P., \& Scoville, N. Z. 2001a,  in preparation

\reference
Jogee, S. Baker, A. J., Sakamoto, K., \& Scoville N. Z. 2001b, \baas, AAS 
Meeting 198, 74.04
 


\reference
Sakamoto, K., Okumura, S.~K., Ishizuki, S., \& Scoville, N.~Z. 1999, ApJ, 525, 691

\reference 
Toomre, A. 1964,  ApJ,  139, 1217 

\end{references}
\end{document}